# MeLM, a generative pretrained language modeling framework that solves forward and inverse mechanics problems

Markus J. Buehler[1,2]*

[1] Laboratory for Atomistic and Molecular Mechanics (LAMM), Massachusetts Institute of Technology, 77 Massachusetts Ave., Cambridge, MA 02139, USA

[2] Center for Computational Science and Engineering, Schwarzman College of Computing, Massachusetts Institute of Technology, 77 Massachusetts Ave., Cambridge, MA 02139, USA

*mbuehler@MIT.EDU

**ABSTRACT**: We report a flexible multi-modal mechanics language model, MeLM, applied to solve various nonlinear forward and inverse problems, that can deal with a set of instructions, numbers and microstructure data. The framework is applied to various examples including bio-inspired hierarchical honeycomb design, carbon nanotube mechanics, and protein unfolding. In spite of the flexible nature of the model–which allows us to easily incorporate diverse materials, scales, and mechanical features–it performs well across disparate forward and inverse tasks. Based on an autoregressive attention-model, MeLM effectively represents a large multi-particle system consisting of hundreds of millions of neurons, where the interaction potentials are discovered through graph-forming self-attention mechanisms that are then used to identify relationships from emergent structures, while taking advantage of synergies discovered in the training data. We show that the model can solve complex degenerate mechanics design problems and determine novel material architectures across a range of hierarchical levels, providing an avenue for materials discovery and analysis. Looking beyond the demonstrations reported in this paper, we discuss other opportunities in applied mechanics and general considerations about the use of large language models in modeling, design, and analysis that can span a broad spectrum of material properties from mechanical, thermal, optical, to electronic.

1. Introduction

The emergence of large language models (LLMs) has challenged conventional strategies of modeling problems across a variety of scientific and engineering disciplines [1–16]. Such models can complement conventional multiscale modeling for the analysis and design of hierarchical materials [17–20], and many other applications in mechanics [21]. Specifically, data-driven modeling using machine learning and related approaches has emerged as a powerful strategy [22–28] that includes both analysis tasks (such as, predicting properties from protein sequences or microstructures) and inverse design tasks (designing novel proteins, biomolecules or composites to meet a set of target properties) [29]. Specifically, generative mechanics is an emerging frontier in materials discovery that complements earlier work in proteins [30] and architected materials [31–34]. Attention-based transformer models [15,35–42] have been shown to display emergent behaviors [43] with important questions that should be investigated specific to applications in science and engineering, and as explored here, applied mechanics. The possibility to realize neural implementations of category theory-based material models is achieved here through the use of graph-forming attention neural networks that construct intricate knowledge graphs from data [41,44,45] through Transformer models [42,46,47]. While many existing models trained on large corpi of data like GPT-3/4, PalM/Bard, or ChatGPT offer impressive general capabilities, they often lack specific domain knowledge and the ability to solve a particular problem (and can, in some instances, hallucinate predictions). We can overcome these challenges by creating a new framework specifically developed to solve domain problems, ultimately forming a sort of "MechGPT" for mechanics. This requires careful development, training, and validation.

The rationale for using attention-based transformer models to solve mechanics problems is rooted in the insight that Transformer architectures provide a first-principles based, flexible framework to capture complex relationships in data, using a language approach that breaks down mechanics into elementary building blocks. Other neural architectures (e.g. fully connected deep neural nets, convolutional operators, graph neural networks, recurrent models like RNN or LSTM, etc.), while successfully applied to specific problems, can not easily generalize beyond the core application against which they have been trained for. In convolutional models (e.g. CNN) filters are applied in Euclidian space, focusing on *neighboring* pixels or voxels. By successively applying CNN layers we can effectively achieve a multi-level coarse-graining strategy akin to physics-based strategies. However, this approach can be limiting especially when dealing with complex data such as genes, language or equations, sequences, or time-series, because long-range relationships cannot be learned on fine-grained details. To generalize this concept, graph neural nets (GNNs) have been widely used for mechanics problems, molecules, proteins, and composites. GNNs relax the constraint operators applied in Euclidian space and expand it to treat apply convolutional operators in the space of adjacencies defined by the graph's edges and associated features (e.g., node features may include the atom type, and edge features the bond length, or bond order). While these models often perform very well, they typically require graphs to be known. Transformer models can be seen as architectural frameworks that generalize even further – to apply operators in graph space, while also discovering the graph itself. In addition, they possess a heightened capacity to discover complex operators beyond conventional convolutional filters, including the possibility for long-range processing. This offers a much more general formulation, following from the concept of neural ologs [41,48], and motivates a theoretical basis for their high expressivity, generalization capacity and emergence potential. In this vein, deep Transformer models effectively represent a large multi-particle system – consisting of hundreds of millions of particles (neurons), where the interaction potential is defined by the nature of the graphs formed, and scaled up through a large set of feed forward layers and head layers during which the interaction potentials discover function from emergent structures [49]. Since Transformers learn to generalize operators across all data seen, and since they are trained to generate sparse graphs via the use of softmax functions (see equations (1-3) in **Materials and Methods**), they capture and generalize knowledge well and are able to discover, and draw upon, synergistic relationships between isolated patterns or sources of data. This may explain why Transformers excel when they are large, and trained on increasingly complex tasks.

In this paper we focus several mechanics problems to illustrate the method as a platform approach (bio-inspired hierarchical honeycomb design, carbon nanotube mechanics, and protein unfolding mechanics, giving three examples in different material domains). The applications include broadly relevant tasks to predict the constitutive stress-strain response for a given microstructure, and to design microstructure candidates so that it has a particular required stress-strain response. **Figure 1** shows an overview of the problems solved in this study. **Figure 1a** depicts the two main problems modeled, including the forward problem (predict constitutive response from a microstructure) and the inverse problem (predict candidate microstructures to meet a specified stress-strain response). **Figure 1b** depicts details on the honeycomb design, inspired by natural honeybee hives, and potential additive manufacturing routes (these are not explored in this study). The right panel in **Figure 1c** depicts the hierarchical design principles underlying these structures, as first reported in [50]. **Figure 1c** shows the stress-strain response calculated using coarse-grained molecular mechanics modeling (these results, including details, are contained in earlier work [50]; here we use the dataset generated therein to develop, train and test the language model). The constitutive response to compressive loading is considered in this work, albeit there is no fundamental limitation as to what properties can be described. **Figure 1d** shows a cycle of the forward and inverse problems for an example microstructure, setting the stage for the problem tackled in this paper. **Figure 1e-f** show additional applications studied in this paper, including carbon nanotube nanomechanics and protein unfolding mechanics, illustrating the boundary conditions of the mechanical tests performed (tensile loading of nanotubes to determine its modulus, and unfolding of a protein by pulling apart its ends while recording the maximum resistance force).

The plan of this paper is as follows. First, we introduce the overall approach of using language models to address mechanics problems, including model development and validation. We then report a series of applications studies including the design new hierarchical composite structures with targeted mechanical properties, modeling and design of nanotube mechanics, and protein unfolding. We focus on a careful analysis of the accuracy of the predictions and the suitability to process applied mechanics problems that have unique or degenerate solutions. We conclude with an outlook to challenges and opportunities.

## 2. Results and Discussion

This work uses multi-headed attention mechanisms that are used widely in many transformer models for sequence and graph data [51]. One of the appeals of this architecture is that it can easily be trained, allows for pre-training and fine-tuning strategies, and can capture extremely complex linguistic, scientific and mathematical relationships [52,53]. Here we show how a LLM framework can be used to accurately solve mechanics tasks.

**Figure 2** shows a general flowchart overview of the strategy implemented in this study. **Figure 2a** provides details of the model is used to solve a variety of problems, such as calculate properties or generate new designs. The input is provided to the model as utf8 encoded string data, which includes tasks, numbers, and other details of the problem solved, then provided to the model, which then samples probability distributions of the output text. The model thereby predicts text in response to the queries, which can be mined to either extract numerical values (e.g. resulting stress-strain response) or a microstructure design. **Figure 2b** shows two complete sample prompts. The model is easy to train. This is because during training, the entire sequence of 'ground truth' is fed, consisting of the task, and the result (marked in **bold**). A trained model can easily be extended to learn new tasks (e.g. add protein mechanics to its base knowledge about honeycomb mechanics), or fine-tuned with new or more accurate data.

When the model is used to make predictions, that is, during inference, only the task is provided, and the model predicts the result of the task by *completing* the sentence (that is, predicting the **bold** characters). The model can be trained to feature several additional, related or unrelated tasks. Sampling can be repeated multiple times and the resulting predictions (e.g. material designs) can be tested against desired outputs or other parameters to create a set of possible designs that can be screened for performance. If unique solutions exist, repeated sampling results in consistent responses. If multiple solutions exist, repeated sampling results in answers that represent the distribution of solutions (e.g. multiple designs, or a distribution of values).

### *2.1 Hierarchical honeycomb mechanics*

We first discuss our experiments conducted with hierarchical honeycomb materials (it is noted, however, that the method can be adopted easily for other microstructure data). Two methods are presented, one specific to the type of hierarchical honeycomb design based on underlying mathematical relations, expressed via a gene vector [50], and a more general discreate, quantized autoencoding strategy that can be applied to any type of high-dimensional data for use in MeLM. The reason for reducing the dimensionality of microstructure data is that it is not efficient to directly encode images, for instance, as text format (since the sequences would be extremely long, and hence render the model highly memory intensive as the matrix multiplications in building the interaction graphs generally scales nonlinearly with the length of the input sequence). **Figure 3** shows details of these two strategies.

The first method (**Figure 3a**), as first reported in [50], uses a gene vector to define a mathematical encoding of the microstructure. In this approach, the hierarchical honeycomb microstructure is defined by a series of numbers, e.g.:

```
12, 4, 7, 3, 0, 0, 0, 0, 0, 0
```

denoting the gene vector. Then, the property calculation task is defined as seeking a calculation on such a gene vector:

```
GeneGetProperties< 12, 4, 7, 3, 0, 0, 0, 0, 0, 0> [-0.988,-0.913,-0.835,-0.760,-
0.695,-0.629,-0.567,-0.520,-0.517,-0.511,-0.503,-0.497,-0.489,-0.481,-0.475,-0.467,-
0.461,-0.452,-0.445,-0.435,-0.430,-0.420,-0.410,-0.402,-0.393,-0.384,-0.374,-0.364,-
0.352,-0.339,-0.320,-0.300]
```

Similarly, the inverse task is given by:

```
GeneGetStructure<-0.987,-0.917,-0.863,-0.823,-0.781,-0.744,-0.702,-0.653,-0.623,-
0.612,-0.611,-0.606,-0.595,-0.584,-0.572,-0.560,-0.550,-0.540,-0.531,-0.520,-0.510,-
0.500,-0.487,-0.469,-0.446,-0.416,-0.386,-0.348,-0.302,-0.260,-0.236,-0.208> [ 8, 8,
8, 8, 8, 0, 0, 0, 0, 0]
```

While the study of hierarchical honeycombs was facilitated via structural specification through gene vectors as developed in [50], this strategy is not generally applicable. For instance, many engineering structural designs cannot be captured by parametric equations (or are unknown *a priori*), and biological designs (e.g. leaf microstructures) do not follow any known mathematical laws. To address this generalization challenge, **Figure 3b** outlines a more flexible approach. Following the vector quantized variational autoencoder model (VQ-VAE) reported in [54] we use a codebook language-based representation of microstructures (therein, a microstructure image is encoded via 64 tokens defined by integer numbers between 0…127 in order to realize an encoded language representation (it is noted that the VQ-VAE model itself uses an attention-based architecture mixed with convolutional operators). An example for one of the hierarchical honeycombs would be:

```
97,121, 43, 43, 20,110,115,  2, 43,110, 22,110, 87, 43,121, 36,  2, 97,
87,113,110,113, 47,121,115,113,110,113, 26,121, 27, 20, 87, 87,121,121,110,121,121,
43,110, 27, 87, 27, 22,110, 47,113, 43, 27, 99,121, 87, 20,121, 43,115, 97,
36,113,121, 20, 47, 20
```

Conceptually this is similar to the gene vector approach but the particular encoding is discovered by the vector quantized variational autoencoder model, and can be identified easily for a diverse range of high-dimensional data. Then, similar as before the tasks are given by, starting with the forward task:

```
GetProperties< 97,121, 43, 43, 43,121, 97, 26, 43, 62, 97, 26, 87,121, 87, 43, 47, 47,
87, 20, 26,121, 20,113, 87,110, 20,121,  2, 47, 26, 26,  2,110,121,121, 87, 87, 97,
 2, 43, 47,  2, 26, 47,121,110, 20, 47, 47, 87, 22,110,121, 20, 97, 47, 87,113,121, 26,
62, 87, 99> [-0.987,-0.928,-0.865,-0.849,-0.840,-0.819,-0.799,-0.779,-0.747,-0.732,-
0.707,-0.693,-0.665,-0.624,-0.589,-0.550,-0.544,-0.496,-0.466,-0.432,-0.411,-0.384,-
0.348,-0.331,-0.295,-0.252,-0.200,-0.150,-0.087,-0.013,0.042,0.098]
```

And for the inverse task:

```
GetStructure<-0.987,-0.928,-0.865,-0.849,-0.840,-0.819,-0.799,-0.779,-0.747,-0.732,-
0.707,-0.693,-0.665,-0.624,-0.589,-0.550,-0.544,-0.496,-0.466,-0.432,-0.411,-0.384,-
0.348,-0.331,-0.295,-0.252,-0.200,-0.150,-0.087,-0.013,0.042,0.098> [ 97,121, 43, 43,
43,121, 97, 26, 43, 62, 97, 26, 87,121, 87, 43, 47, 47, 87, 20, 26,121, 20,113,
87,110, 20,121,  2, 47, 26, 26,  2,110,121,121, 87, 87, 97,  2, 43, 47,  2, 26,
47,121,110, 20, 47, 47, 87, 22,110,121, 20, 97, 47, 87,113,121, 26, 62, 87, 99]
```

All training data, no matter which description of microstructures (or other data, see sections below), properties, or features is used, is fed into an autoregressive transformer architecture, which processes text using a byte-level tokenizer that can effectively and flexibly deal with a variety of string data in utf8 encoding (**Figure 4**, see **Materials and Methods** for details). Both numbers and text characters (various symbols, including math symbols, LaTeX, etc.), allowing also for human interaction with the model as an interactive modality, is possible. No special encoding is necessary for numerical input and output, providing a straightforward way to build datasets for training.

Once the model is trained, we can use it to test its performance on a variety of cases. We start by looking at the overall performance of the forward and inverse tasks, and then dive deeper into specific examples. **Figure 5** shows results from forward and inverse tasks, depicting how the predicted stresses compare with the ground truth (GT) stresses. **Figure 5a-b** show the result for the model using gene vector microstructure encoding (as shown in **Figure 3a**; **Figure 5a**, forward task: R2=0.95, **Figure 5b**, inverse task: R2=0.96), and **Figure 5c-d** show the result for the vector quantized language encoding (**Figure 5c**, forward task: R2=0.96, **Figure 5d**, inverse task: R2=0.91). The two microstructure encoding strategies show similar performance, albeit the gene vector approach yields slightly better performance for both forward and inverse tasks (this can probably be addressed with more training data). The sampling temperature for all predictions is $T_{sample}$=0.01.

Using the gene vector encoding strategy, **Figure 6** shows results from solving the inverse task and comparing the constitutive behavior of the designs with the desired response (as shown in **Figure 4c**, top panel). The results are obtained for a test set of microstructures not included in the training of the model. For the three examples shown, it can be seen that the model performs well in matching the desired stress-strain response. The right column, labeled GT, shows the microstructure designs in the dataset. As can be seen, the model proposes new, distinct, designs that yield the desired response. Similarly, for the vector quantized language encoding strategy, **Figure 7** shows results of solving the inverse task and comparing the constitutive behavior of the designs with the desired response. The model does an excellent job in matching the desired stress-strain response. As before the column labeled GT depicts the microstructure designs in the test dataset. As can be verified in the figures, the model proposes new, distinct, designs that yield the desired response. **Figure 8** shows an overview of solutions to the forward problem, for the gene vector encoding strategy (**Figure 8a**) and the vector quantized language encoding strategy (**Figure 8b**). For all cases, the model can effectively and accurately predict stress-strain responses for a given microstructure. The sampling temperature in all cases is $T_{sample}$=0.01. One of the special features of the model is that it because it has learned to solve both the forward and inverse problem, it can be immediately used to check whether or not solutions to the inverse problem are cycle consistent. This is done by feeding the prediction of the inverse model into the forward model and then comparing the desired response and the measured responses. Further validations could be performed, like using experimental essays, or physics-based modeling. These experiments were done, and the results visualized, in the analysis shown in **Figure 5b and d**, as well as **Figures 6-7**

One of the exciting opportunities of the model is that it can innately capture the development of degenerate solutions, if they so exist. As shown in **Figure 9**, by conducting predictions multiple times, the model can predict multiple solutions to the same task. To explore this scenario we show examples of three different microstructure designs that all meet the same design target (using the gene vector encoding strategy). All predicted designs feature the same hierarchical honeycomb patterns, but offer a great level of diverse designs. Further analysis, such as assessing density or material use, or which model fits certain regions in the desired stress-strain response better, can be used to identify the best candidates for a selected microstructure. Increased variation of the design can be obtained by increasing the sampling temperature, in ranges from $T_{sample}$=0.1 to 1 (however, as the temperature is increased, the constraint of the target solution is more and more relaxed, leading to solutions that deviate more strongly from the design objective as the model becomes more creative). We note that the sampling temperature for the forward problem is always set to $T_{sample}$=0.01 to ensure the highest-accuracy model predictions for this task. Moreover, for problems where only one solution exists – e.g. a given microstructure's response as defined in the forward task, repeated sampling will not lead to multiple solutions but yields the same response. Increased sampling temperature merely leads to higher deviation from the baseline solution.

Now we move on to experiments conducted with the vector quantized language encoding of microstructures, shown in **Figure 10**. It can be seen that these *de novo* designs feature a greater variety of solutions that clearly exceed those of the original training data. This is because its language-based design language captured by the vector quantized autoencoder model is not limited to the mathematical operations behind the gene vector encoding approach. Hence, this model can identify designs that are distinct from the architectural families of geometries included in the training set. While the two left designs closely follow the patterns in the training set, the model can also produce new designs, as shown in the right examples. Generally, increased variation of the design can be obtained by increasing the sampling temperature (as before, the sampling temperature for the forward problem is always set to $T_{sample}$=0.01).

## 2.2 Incorporating carbon nanotube and protein mechanics

Next, we explore whether a trained model (here, the gene encoding model) can be expanded, with additional data, to additionally capture other mechanics applications. We consider two examples, first predicting elastic properties of carbon nanotubes and designing carbon nanotube candidates to meet a certain modulus, and second, to predict the unfolding strength of a protein directly from its sequence without the need to know a 3D structure. The tasks used for the nanotube application are

```
GetCNTModulus< 41,  5> [0.518]

GetCNTStructure<0.558> [ 44,  0]
```

As before, the prompts consist of the task and the predicted result (in bold) – the first one, to calculate the CNT modulus from chiralities, and the second one to get CNT chiralities that meet a modulus design demand. **Figure 11** shows results of this experiment. As the plots show, the model displays excellent performance in predicting the modulus and generating CNT designs that meet the design objective. Notably, even though we added a distinct set of tasks to the model, it still retains an excellent ability to design hierarchical honeycomb structures. This points to exciting potential to train this model with a larger number of mechanics tasks, scaling it to incorporate larger corpuses of data. However, we believe it is important to carefully examine the successive development of such models as a foundation.

We further explore whether an even more complex task such as end-to-end protein property prediction can be added. **Figure 12** shows the results, revealing that the model can accurately predict the unfolding strength of protein sequences. For this case, the prompt is:

```
GetProteinForce<MSYYHHHHHHLESTSLYKKAGSENLYFQGNKKDIPWTDLNRASGVGSTGILQAR
IINGVIYVRGNSIPVPNVAPNFIVPVGTFPPAFGTNLPQFDSSGTFYSHGNLSLSLINMSPSGIAVGNPN
NTSMNGKTISFALSAPLL> [0.365]
```

Two proteins with extreme properties are highlighted **Figure 12c**. In agreement with understanding of protein mechanics, the strongest protein (depicted as red square) is highly organized and rich in beta-sheet secondary structure (reaching a strength of ~ 376.68 pN). Conversely, the weaker protein (shown as a blue circle) reaches a strength of just 74.1 pN. The relative weakness of this protein can be explained by the structure, showing more disordered regions with only small helical and sheet domains. To show that the model is still capable of solving hierarchical honeycomb design tasks (in alignment with the excellent R2 values seen in **Figure 12d-e**), we depict a sample design task in **Figure 12f**. It is anticipated that the protein tasks could be developed towards more accurate predictions, e.g. using a pre-training strategy (unsupervised training of large sets of protein sequences, which is likely advantageous for highly complex problems like protein behavior). We did not yet train an inverse task for protein design, and leave this to future work (as it would require folding the new sequences, followed by molecular simulations, to adequately test the predictions). Generally, we anticipate that such tasks can be accomplished well as evidenced by recent studies. This is left to future work.

### 3. Conclusions

We proposed MelM, a mechanics language model based on autoregressive, decoder-only transformer architectures, which provides a flexible platform for diverse mechanics problems, for both forward and inverse problems (see, **Figures 1-4** for problem definition and overall approach). The flexible nature of the modeling strategy, applied here to systems ranging from hierarchical honeycombs to carbon nanotubes to protein mechanics allows us to easily incorporate these models into a wide range of applications and solve multiple and complex tasks, as shown in **Table 1**. While the studies reported here were focused on only three sample applications, we anticipate that it can be generalized towards other use cases with the potentiality of emergent behavior across tasks, modalities (mixing text, images, numbers, etc.), and domains of knowledge [3,30,32,55]. Especially via the use of the VQ-VAE [54,56] model to encode diverse high-dimensional data in language form, numerous applications can be realized.

The model was trained in stages, showing great potential for adding further data, knowledge or variations of the tasks as data or needs grow. It is also conceivable to train a model on a larger set of unlabeled data (e.g. more microstructures encoded by the VQ-VAE model) and then fine-tune on a smaller, labeled dataset. Such a strategy can, for instance, provide a reservoir of design ideas that can be used to solve inverse problems, where the solutions can exceed those of the training data. Similar emergent behavior has been seen in related applications in chemistry [14] and protein properties [3]. The use of the model in the context of ingesting molecular modeling results and generalizing these towards rapid, effective and generative predictions (as done for all three application examples) is a powerful upscaling strategy that can complement existing multi-scale methods.

It is anticipated that this class of models can find many useful applications in applied mechanics and related fields (e.g. fracture, dislocation mechanics, diffusion, dynamics, etc. [57–64]), especially due to its ease of training and use. One of the compelling features is the lack of a need to encode numerical values differently from text. Here, numbers and their representations (*i.e.* characters 0…9 and punctuation, like ',' are simply part of the elementary level of building blocks and form part of the input, output, and are naturally part of the interactive dialog with the model). This implies that such a model could easily be trained on mechanics theory, proofs, derivations, and other concepts used widely in theoretical and applied mechanics. By ingesting textbooks (e.g. equations formatted consistently in LaTeX), this theoretical framework may be able to learn powerful concepts that can be used in the development of new insights or solutions to problems. Within this context, a particularly exciting avenue is to use pre-trained models such as GPT-4, PaLM, LaMDA, or Falcon. The advantage of using models trained on diverse knowledge is that they generate additional convergent insights that can collectively lead to emergent behavior. This potentiality can be tapped into by fine-tuning these to incorporate both linguistic and logical reasoning capabilities (these exist in such general-purpose models) with the ability to solve mechanics problems (the focus on the MeLM model) [5,6,65,66]. The models reported here, and other large language models, can be effectively fine-tuned by using strategies such as Low-Rank Adaptation of Large Language Models (LoRA) adaptor methods [67]. Such strategies are particularly effective to avoid catastrophic forgetting, which can complement fine-tuning tasks that need to rely on earlier training data in downstream deployments.

## 4. Materials and Methods

### 4.1 Dataset construction and tokenizer

We use byte-level tokenization similar to that used in the T5 model [68] to represent UCS Transformation Format 8 (utf8) codes in 384 tokens (each character is represented by one to four bytes; the vocabulary size includes the 256 utf8 byte-level tokens plus a series of special tokens that are not yet used here, but can be used in future adaptations of the model). We thereby encode a variety of tasks including a mix of numbers and text characters; and it also opens the door to future training/fine-tuning of the model to include other datasets. Token sequences are encoded using trainable embedding layers. Various examples of the string construction are given in the main text and in **Table 1**.

**Figure 1c** shows the stress-strain response calculated using coarse-grained molecular modeling used to construct the dataset. These results are contained in earlier work (see [50]), resulting in a dataset of 1,445 pairs of microstructure and stress-strain data (for the training data, we use these pairings to generate training text data for both forward and inverse tasks, doubling the size). To train the model to understand the design principles of the language encoding based on the VQ-VAE model, we include a Structure<..> training task, where we include all known design sequences as a reservoir for training. If separate pre-training would be used, this task would be trained first, in isolation, as it does not require any labeled data (this is akin to web scraping before fine-tuning a LLM model on downstream tasks, such as translation or question-answering is done).

The carbon nanotube training data is taken from [69], consisting of 818 pairs of chirality (*n,m*) and associated mechanical properties (here, we solely focus on the Young's modulus). In the original work, CNT properties were calculated using molecular dynamics modeling. The protein data is taken from the Bio-molecule Stretching Database (BSDB) [70,71], providing pairs of sequences and associated mechanical properties (here we solely focus on the maximum unfolding force). We clean the data by considering only forces below 475 pN, and lengths below 256 amino acids. A total of 16,618 pairs of data is included. In the original work [71], the unfolding forces were calculated using coarse-grained molecular modeling.

90% of the data is used for training, and 10% as test set. Additional validation cases are considered for the hierarchical honeycomb analysis.

### 4.2 Attention based transformer model

**Figure 4** shows an overview of the transformer architecture used in this study (following an architecture similar to those used in common GPT models [1–3,43]), which processes input tasks defined by strings of text, and predicts outputs from it via a series of causal self-attention layers [42]. We use a decoder-only transformer architecture as shown in **Figure 4a**, featuring rotary position embeddings (RoPE) in order to realize explicit relative position dependency of input tokens as proposed in [72]Rotary position embeddings facilitate broad generalization to varied sequence lengths [73]. The model is developed using causal multi-headed attention to yield an autoregressive model that predicts the next token from the previous ones (**Figure 4b**).

During training, categorical cross-entropy loss is used to train the model to predict the correct next token for training data, where the training data features complete tasks and the resulting output (see **Table 1**). The training strategy can be realized simply by shifting the tokens of the output (forming labels) to the left; hence, the model is trained to predict the next tokens from the previous ones. The model is trained simultaneously to solve the forward and inverse problem. As shown in **Figure 4c**, a trained model can be used to check whether a predicted microstructure yields the desired stress-strain response, to assess cycle consistency.

The key mathematical operation is the masked attention mechanism[11,42]:

$$\text{Attention}(Q, K, V; M) = \text{softmax}\left(\frac{QK^T + M}{\sqrt{d_k}}\right)V \qquad (1)$$

with a triangular mask $M$ (e.g. here for a sequence length 3):

$$M = \begin{pmatrix} 0 & -\infty & -\infty \\ 0 & 0 & -\infty \\ 0 & 0 & 0 \end{pmatrix} \qquad (2)$$

(this masking strategy ensures that the model only attends to tokens to the left to enforce causality. The causal attention calculation is implemented in multi-headed form by using parallelly stacked attention layers, with a total dimension $d$. The softmax function in (1) turns a vector of $L$ real values into a vector of $L$ real values that sum to 1. Instead of only computing the attention once, in the multi-head strategy (where $h$ denotes the number of attention heads) we divide the input into segments (in the dimension of the hidden dimension, that is, $d_{v,i} = d/h$). We calculate scaled dot-product attention over each segment, allowing the model to jointly attend to information from different representation subspaces at different positions:

$$\text{MultiHead}(Q, K, V) = \text{Concat}(\text{head}_1, \ldots, \text{head}_h)W^O \qquad (3a)$$
$$\text{head}_i = \text{Attention}(QW_i^Q, KW_i^K, VW_i^V) \qquad (3b)$$

In self-attention as used in this paper, all $Q$, $K$, $V$ come from either input or output embeddings (or other sources) only.

During generation (inference), the task sequence is fed into the model and the output is predicted from it. During sampling iterations, this process is repeated until the full output is produced. All conditioning and distinction of various tasks is provided by the input prompt. Additional sets of tokens are defined to encapsulate various tasks and input/output boundaries <..> encapsulate task, [..] to encapsulate prediction). Causal autoregressive training is performed using cross-entropy loss, where the next token in the input sequence $z$ is the label for the current token (*i.e.*, labels start with the second token of the input, and we remove the last logit since no label exists, see a visual representation in **Figure 4b**). The training data the task and the corresponding prediction in one sentence, comprising all associated tokens:

$$z = [z_1, z_2, \ldots, z_N] \qquad (4)$$

Gumbel softmax sampling [74,75] is used during inference (however, we also explored the use of beam search). This allows one to adapt the creativity of the model. This is achieved by adding a defined level of noise controlled via the sampling temperature $T$ to a fractional set of logit distributions (identified by a sampling threshold) predicted by the transformer model. We then sample the predicted token from this revised distribution. This helps to add expressivity to the generative tasks to achieve more variations in the predictions ($T$ around or larger than 1). In forward prediction tasks we find that they are best conducted using low sampling temperatures ($T$=0.01).

The model features a dimension of 512 (model used for gene vector microstructure encoding, as well as protein mechanics and carbon nanotube mechanics) and 1024 (model trained for vector quantized encoding). We use a total of 8 attention heads, transformer depth $N_T=24$, and a feed forward multiplier=4 (2048/4096 channels) (126.4M parameters/303.1M parameters, respectively). We use Gaussian Error Linear Unit (GELU) activation functions [76].

*4.3 Vector quantized variational autoencoder (VQ-VAE) model*

The vector quantized variational autoencoder (VQ-VAE) architecture is used as trained in [54]. We provide some background here about this model for clarity and completeness. The VQ-VAE model learns to encode microstructure images (or other high-dimensional data) into a lower-dimensional, *discrete* language space. This discrete codebook representation consists of a one-dimensional vector of length *N* where each entry is one of $n_c$ possible "words" in the design language that defines the microstructures. The encoder and decoder blocks each consist of a deep neural network featuring convolutional and self-attention layers (so that the model can effectively learn both coarse-graining features and long-/and short-range relationships, effectively deciphering a design language that captures the space of microstructures). The VQ-VAE model is trained based on unlabeled data of microstructure images. In further developments, this model can be expanded and trainer either on a different set of microstructures or other data, or to feature additional design elements for greater diversity of predictions that can be introduced in the unsupervised training stage of this model. In its current implementation the model can predict a range of hierarchical honeycomb structures; not limited to the specific designs used during training (as can be seen in the results in **Figure 10**, right two panels), but in the general structural space as the training data).

*4.4 Training process and other hyperparameters*

All code used in this study is developed in PyTorch [77], implemented within the Hugging Face ecosystem. Training is performed using an Adam optimizer [78] with a learning rate of LR=0.0001 and $\varepsilon$=1E-8. We use 1,000 warmup training steps (during which the learning rate is ramped from 0 to the desired learning rate, LR), followed by linear decay (training is performed over 20-30 epochs). During training we assess the R2 value of cycle consistent design tasks during training on a test set, and pick the training epoch with the highest R2 for this task (this ensures good forward and inverse performance). The models are developed in Hugging Face for easy accessibility and further fine-tuning.

**Author contributions**

M.J.B. developed the overall concept and the algorithm, designed the ML model, developed the codes, oversaw the work, and drafted the paper.

**Code availability**: The MeLM model, code, trained weights, and data is available at: https://github.com/lamm-mit/MeLM.

**Acknowledgements**: This work was supported by the MIT-IBM Watson AI Lab, the Army Research Office (W911NF1920098 & W911NF2220213), ONR (N00014-19-1-2375 and N00014-20-1-2189), as well as USDA (2021-69012-35978).

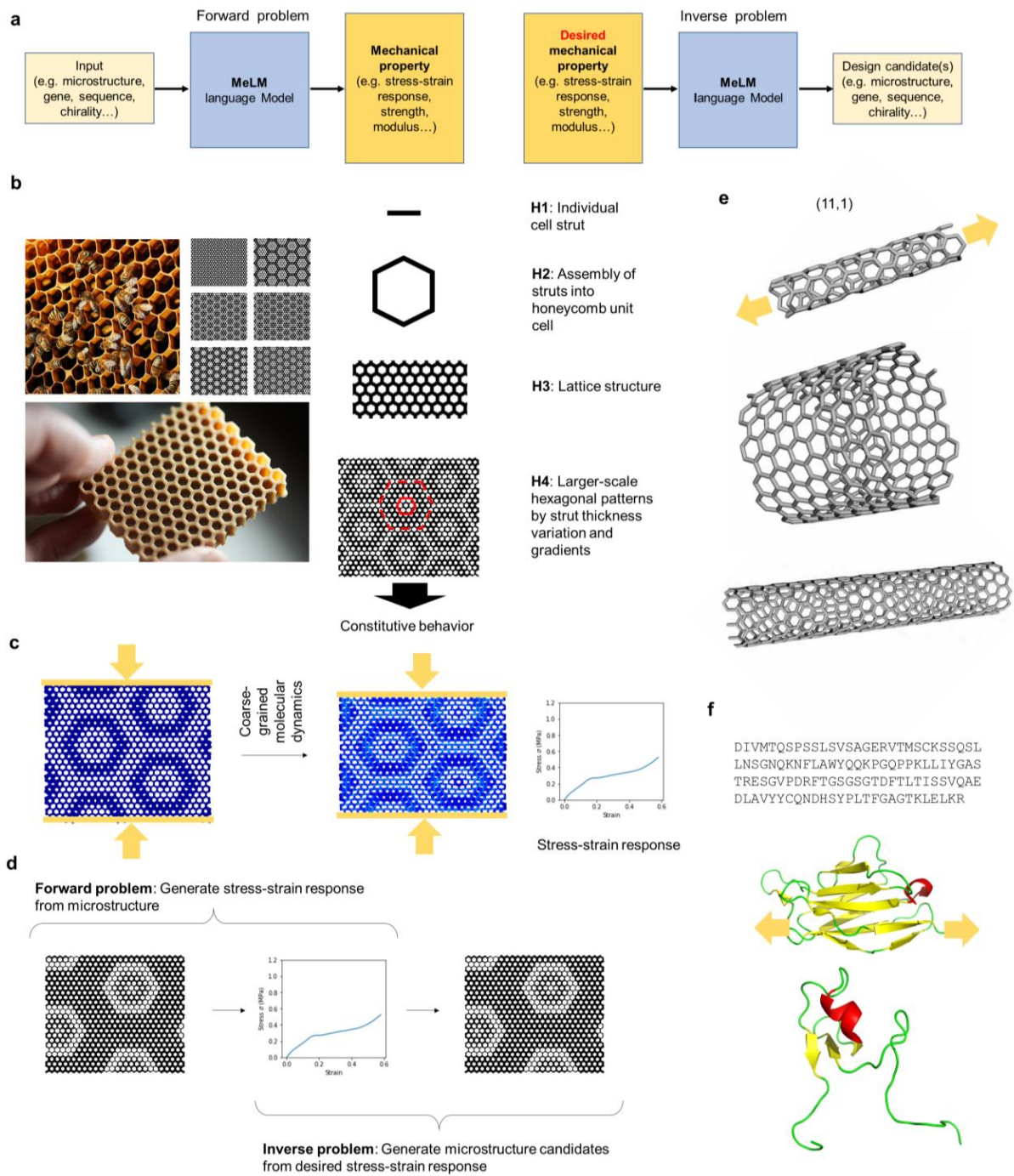

**Figure 1**: Overview of the overall problem solved, and details on the mechanics problem studied in this paper, featuring hierarchical honeycombs, carbon nanotubes and proteins. Panel a shows the two major classes of mechanics problems modeled, including the forward task (e.g., predict constitutive response from a microstructure) and the inverse problem (e.g., predict candidate microstructures to meet a specified stress-strain response). Panel b depicts details on the honeycomb design, inspired by natural honeybee hives (image generated using DALL-E [79]), and potential additive manufacturing routes. The right panel depicts the hierarchical design principles underlying these structures (figure adapted from [54]). Panel d shows the stress-strain response calculated using coarse-grained molecular modeling (as done in earlier work [50], including associated experimental validation). The constitutive response to compressive loading is considered in this work, albeit there is no fundamental limitation as to what properties can be modeled. Panel d shows a cycle of the forward and inverse problems for an example microstructure, setting the stage for the problem tackled in this paper. Panels e and f show additional applications studied in this paper, including carbon nanotube nanomechanics (e) and protein unfolding mechanics (f).

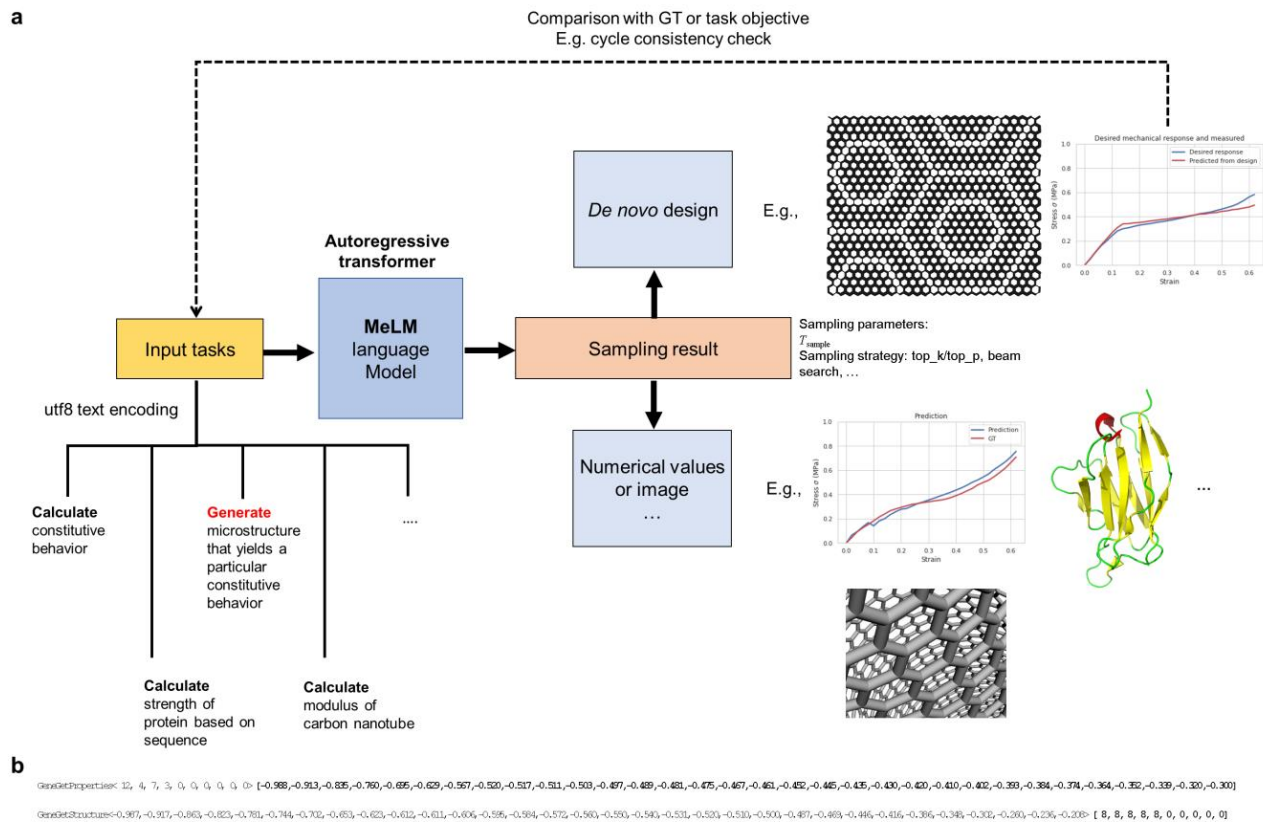

**Figure 2:** Overview of the MeLM model implemented in this study. Panel a: A language model is used to solve a variety of tasks, such as calculate properties or generate new designs. The input is provided to the model as utf8 encoded string data, which includes tasks, numbers, and other details of the problem solved. As ingle language model, implemented as an autoregressive decoder-only transformer model is trained to solve these tasks. The model predicts text output in response to the queries, which can be mined to either extract numerical values (e.g. resulting stress-strain response) or a microstructure design. Panel b shows two sample prompts. During training, the entire sequence is fed (consisting of the task, and the result (marked in **bold**). When the model is used to make predictions (*i.e.* during inference), only the task is provided, and the model predicts the result of the task by completing the sentence (predicting the **bold** characters).

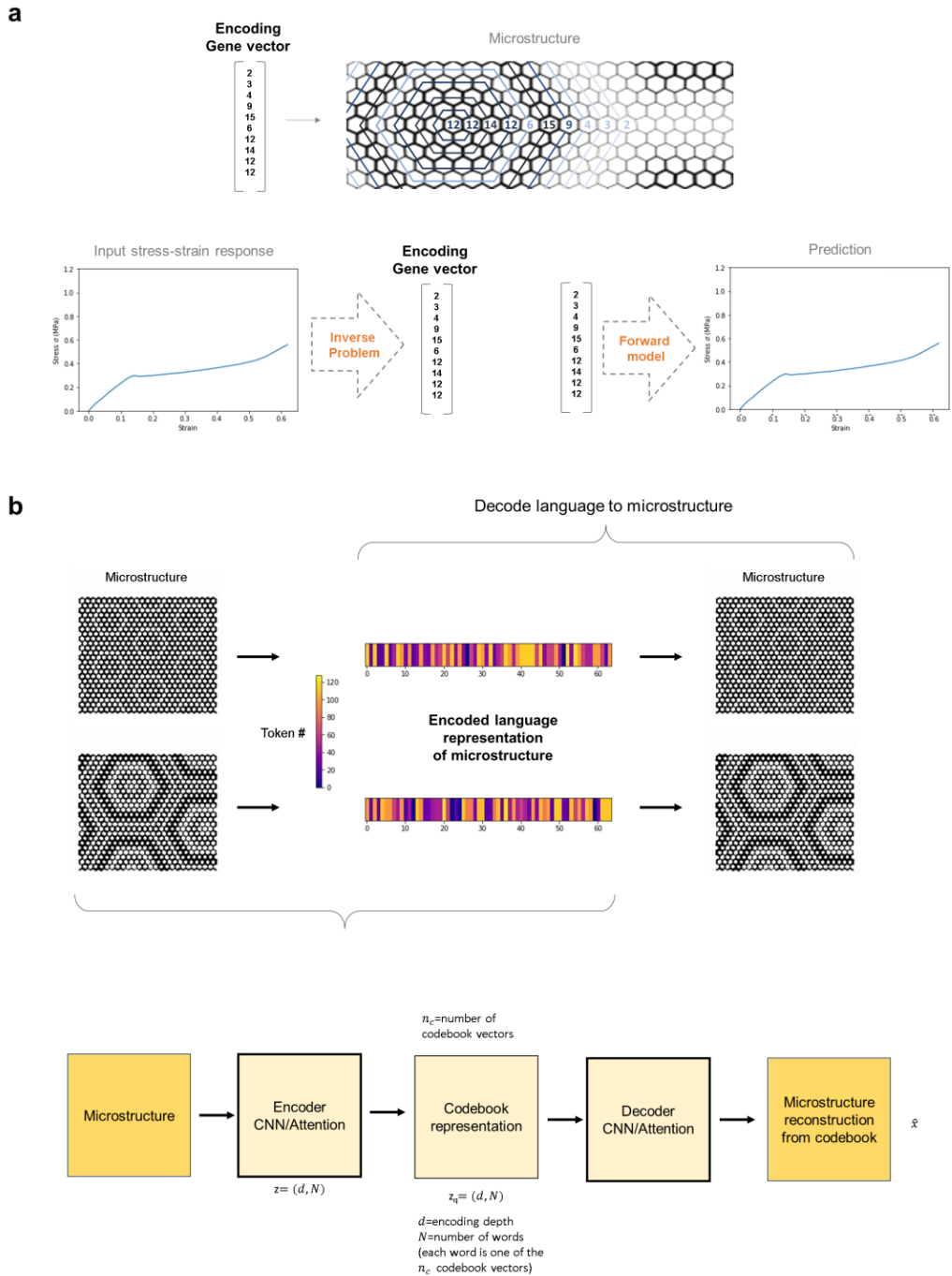

**Figure 3**: Strategies used to encode microstructural design. The first method, as first reported in [50], uses a gene vector to define a mathematical encoding of the microstructure as shown in panel a. In this approach, the microstructure is defined by a series of numbers, denoted the gene vector. Panel b shows a more flexible approach. Following the vector quantized variational autoencoder model reported in [54] we use a codebook language-based representation of microstructures (therein, a microstructure image is encoded via 64 tokens defined by integer numbers between $0\ldots 127$; to realize an encoded language representation). While the gene vector approach in a works very well for the designs studied here, the approach using a flexible language encoding is offers a series of advantages and can be applied to any type of data. This method can be flexibly applied to a variety of other high-dimensional data, such as images, voxel data, etc.

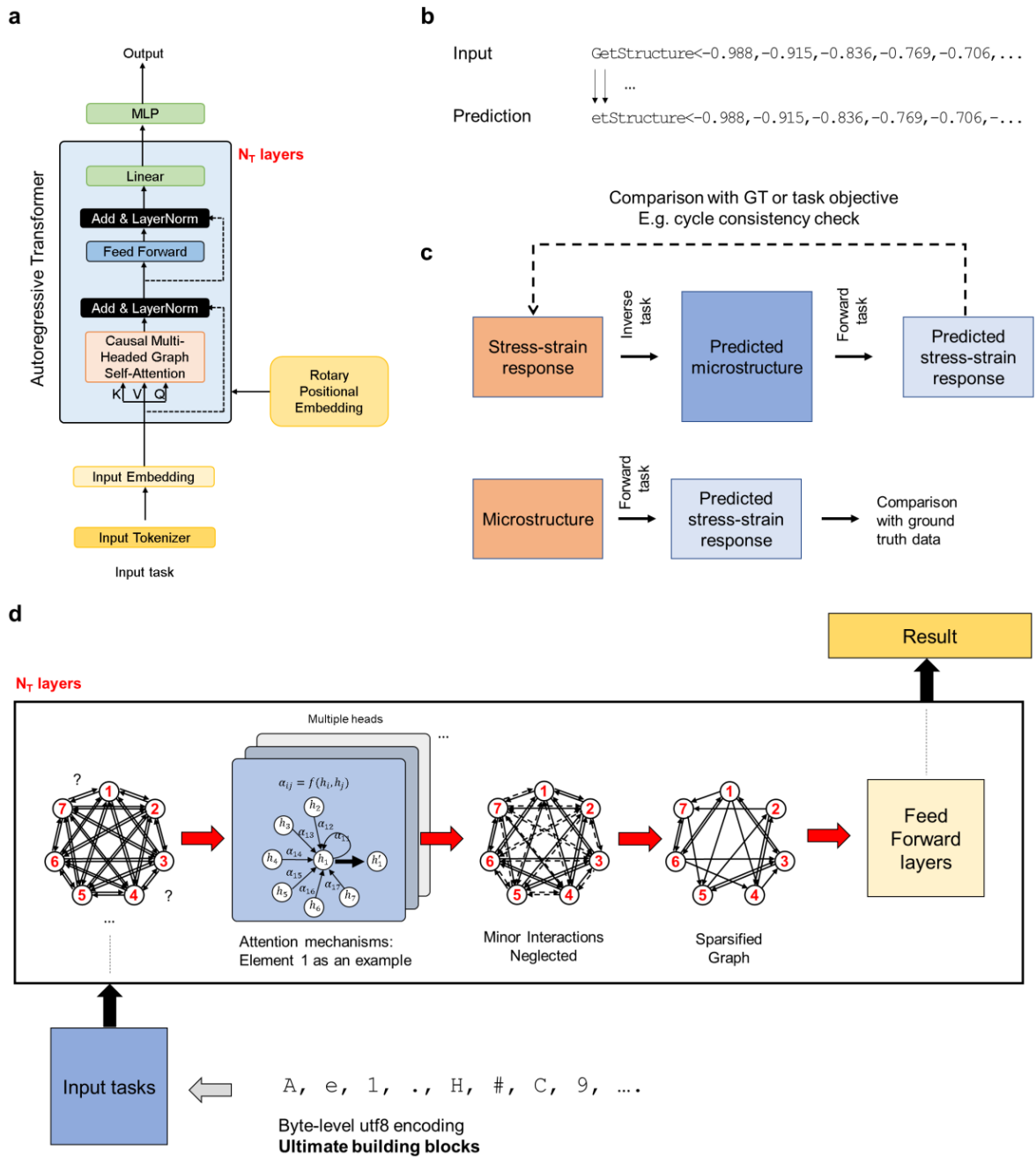

**Figure 4**: Overview of the transformer architecture used in this study (following an architecture similar to those used in common GPT models [1–3,43]), which processes input tasks defined by strings of text, and predicts outputs from it via a series of causal self-attention layers [42]. We use a decoder-only transformer architecture as shown in panel a. The model is implemented using causal multi-headed attention to yield an autoregressive model that predicts the next token from the previous ones (panel b). During training, categorical cross-entropy loss is used to train the model to predict the correct next token for training data, where the training data features complete tasks and the resulting output (see **Table 1**). The training strategy can be realized simply by shifting the tokens of the output (forming labels) to the left; hence, the model is trained to predict the next tokens from the previous ones. The model is trained simultaneously to solve the forward and inverse problem. As shown in panel c, a trained model can be used to check whether a predicted microstructure yields the desired stress-strain response, to assess cycle consistency. Panel d shows an overview of the multi-headed graph forming strategy implemented in the architecture, showing how information is processed across multiple attention heads, resulting in a set of graphs ultimately fed to produce the output probability distribution.

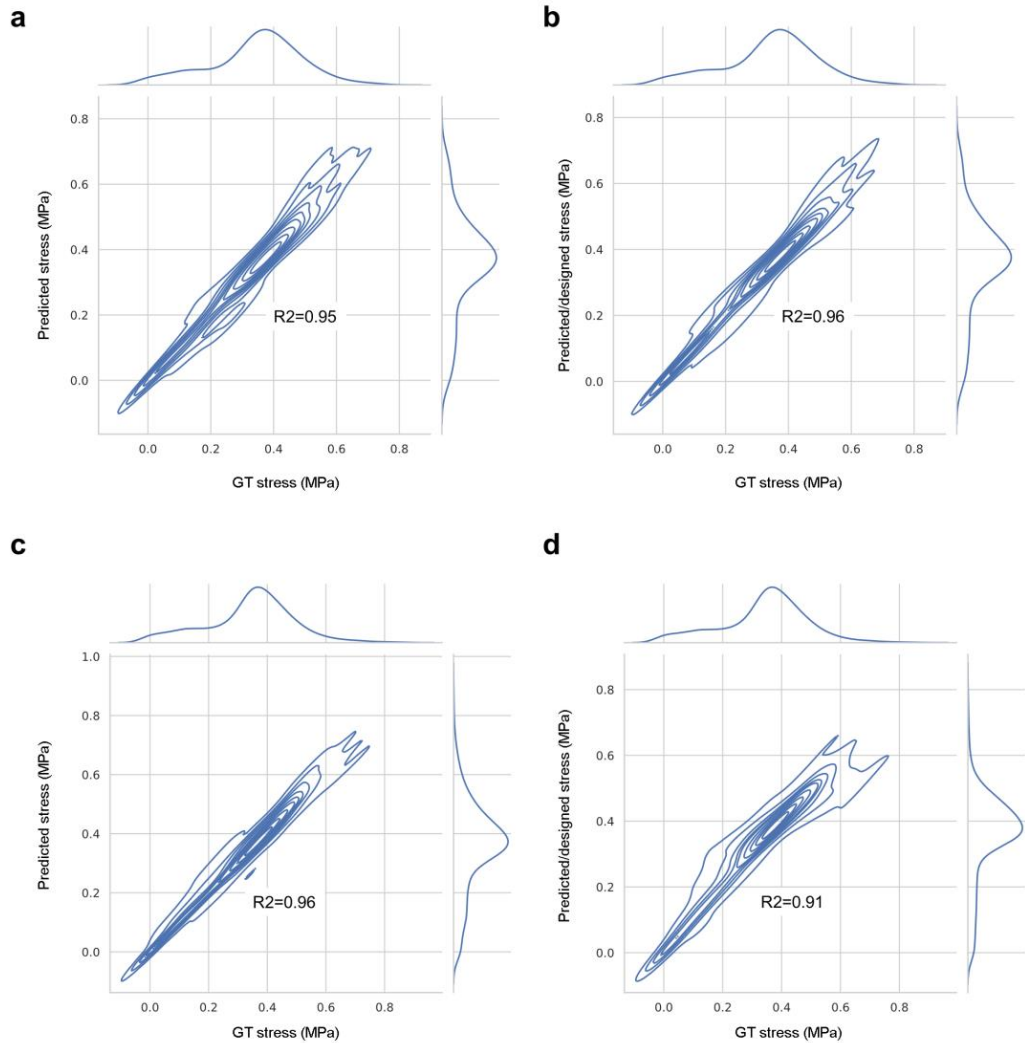

**Figure 5**: Results from forward and inverse tasks of hierarchical honeycomb mechanics, depicting how the predicted stresses compare with the ground truth (GT) stresses. Panels a-b show the result for the model using gene vector microstructure encoding (as shown in **Figure 3a**; a, forward task: R2=0.95, b, inverse task: R2=0.96), and Panels c-d show the result for the vector quantized language encoding (c, forward task: R2=0.96, d, inverse task: R2=0.91). The two microstructure encoding strategies show similar performance, albeit the gene vector approach yields slightly better performance for both forward and inverse tasks. The sampling temperature for all predictions is $T_{\text{sample}}$=0.01.

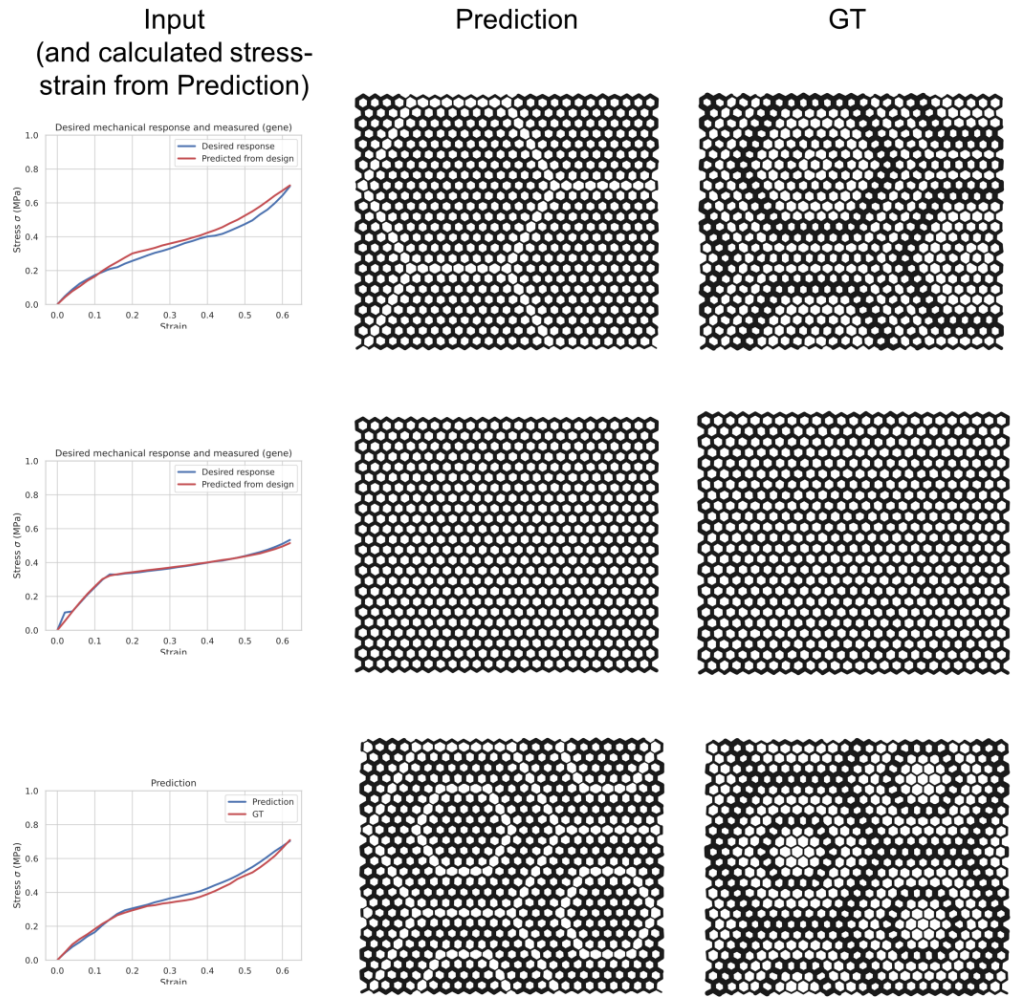

**Figure 6**: Solving the inverse task and comparing the constitutive behavior of the designs with the desired response (as shown in **Figure 4c**, top panel), using the gene vector encoding strategy, for a test set of microstructures not included in the training of the model. For the three examples shown, it can be seen that the model does an excellent job in matching the desired stress-strain response. The right column, labeled GT, shows the microstructure designs in the dataset. As can be seen, the model proposes new, distinct, designs that yield the desired response. The sampling temperature is $T_{sample}=0.01$.

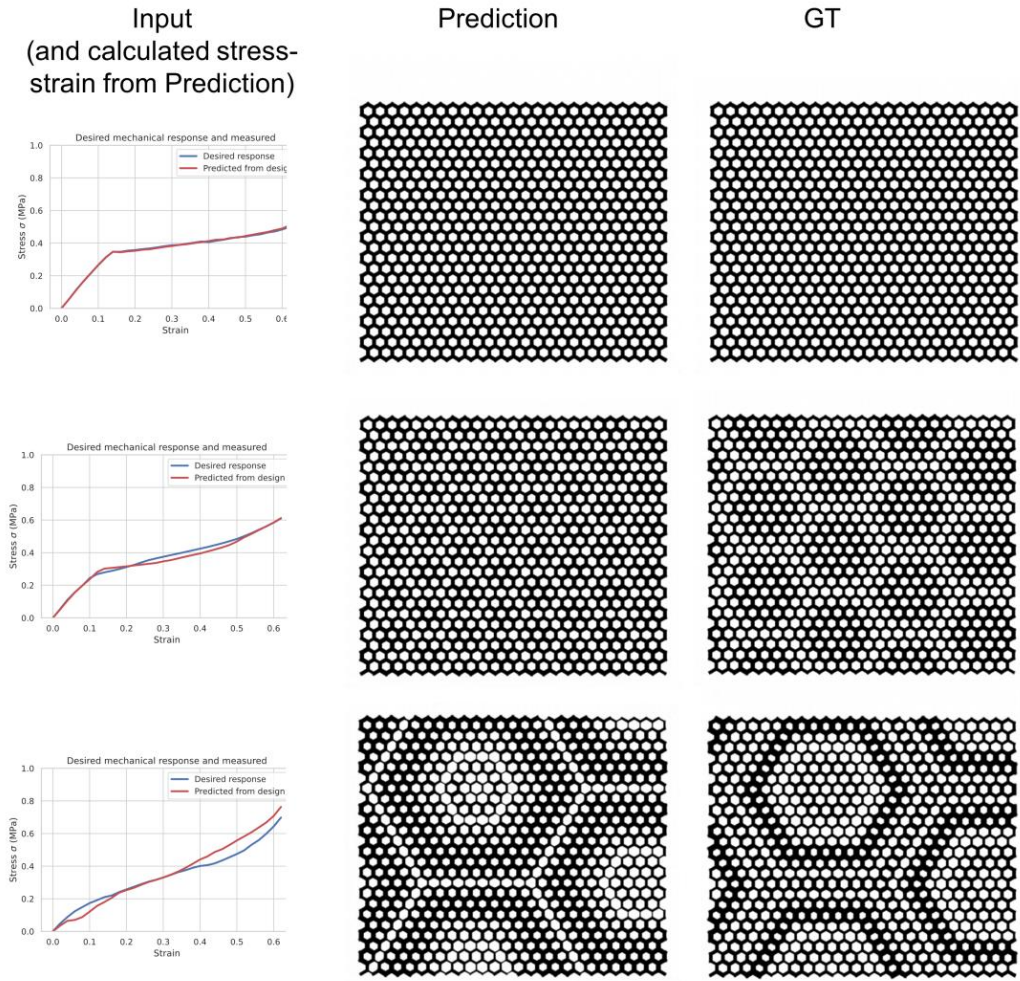

**Figure 7**: Solving the inverse task and comparing the constitutive behavior of the designs with the desired response (as shown in **Figure 4c**, top panel), using the vector quantized language encoding strategy, for a test set of microstructures not included in the training of the model. The model does an excellent job in matching the desired stress-strain response. The right column, labeled GT, shows the microstructure designs in the dataset. As can be seen, the model proposes new, distinct, designs that yield the desired response. The sampling temperature is $T_{\text{sample}}=0.01$.

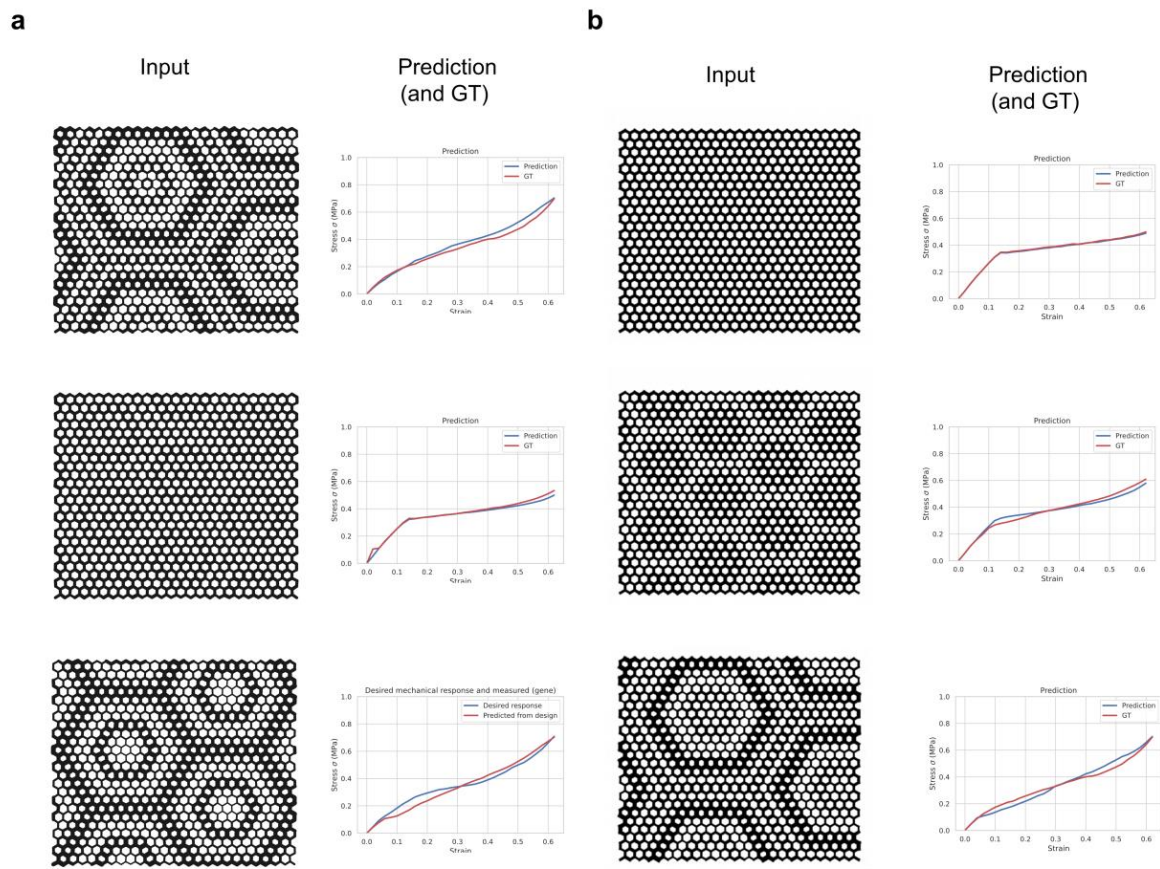

**Figure 8**: Overview of solutions to the forward problem, for the gene vector encoding strategy (a) and the vector quantized language encoding strategy (b). For all cases, the model can effectively and accurately predict stress-strain responses for a given microstructure. The sampling temperature is $T_{sample}$=0.01.

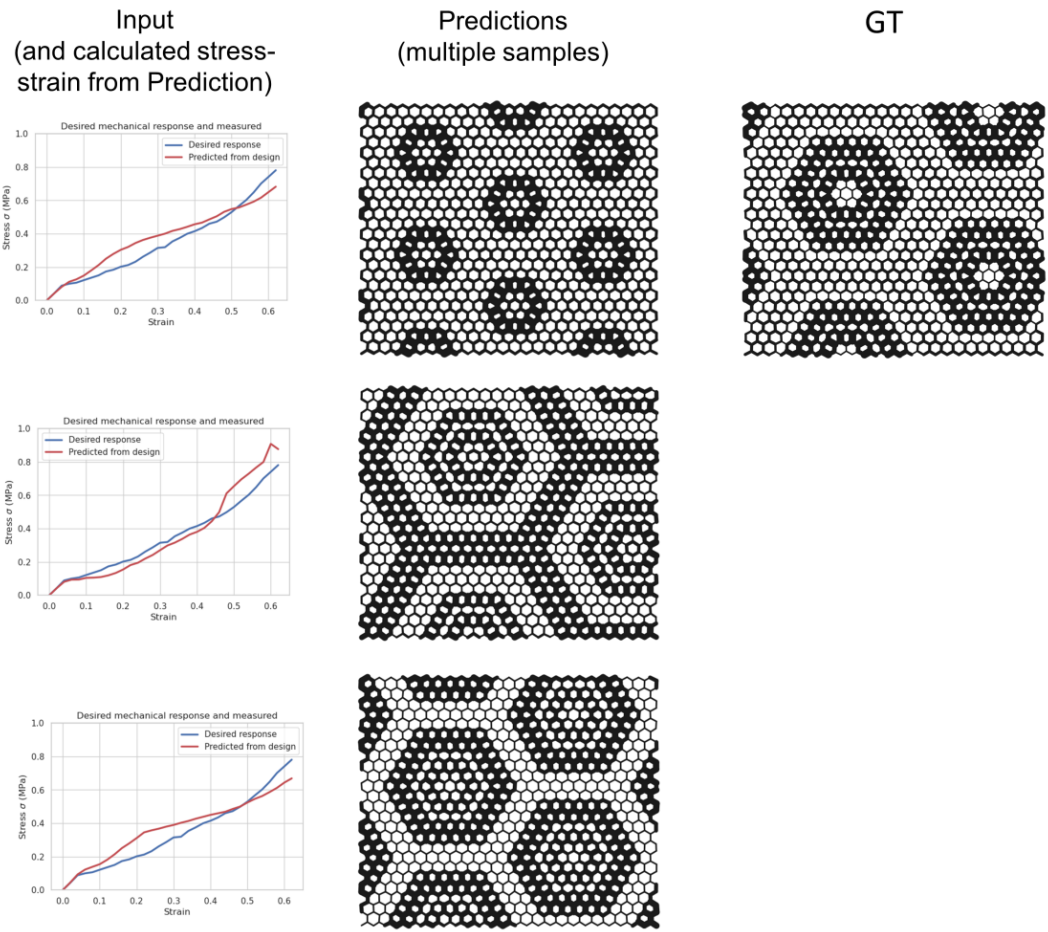

**Figure 9**: By conducting predictions multiple times, the model can predict multiple solutions to degenerate problems. Here we show examples of three different microstructure designs that all meet the same design target. These results were obtained using the gene vector encoding strategy. All predicted designs feature the same hierarchical honeycomb patterns, but offer a great level of diverse designs. Further analysis, such as assessing density or material use, or which model fits certain regions in the desired stress-strain response better, can be used to identify the best candidates for a selected microstructure. Increased variation of the design can be obtained by increasing the sampling temperature. While the sampling temperature for the forward problem is always set to $T_{sample}=0.01$ to ensure the highest-accuracy model predictions for this task.

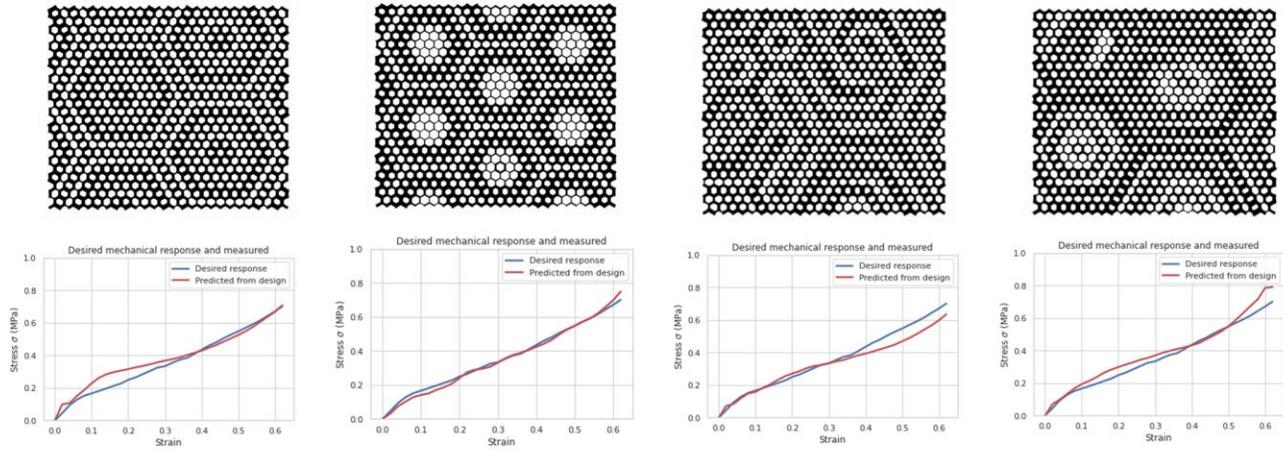

**Figure 10**: *De novo* designs obtained using the vector quantized language encoding. Since the design language is not limited to the mathematical operations behind the gene vector, this model can identify designs that are distinct from the architectural families of geometries included in the training set. While the two left designs closely follow the patterns in the training set, the model can also produce new designs, as shown in the right examples. Generally, increased variation of the design can be obtained by increasing the sampling temperature (sampling temperature for the forward problem is always set to $T_{\text{sample}}=0.01$).

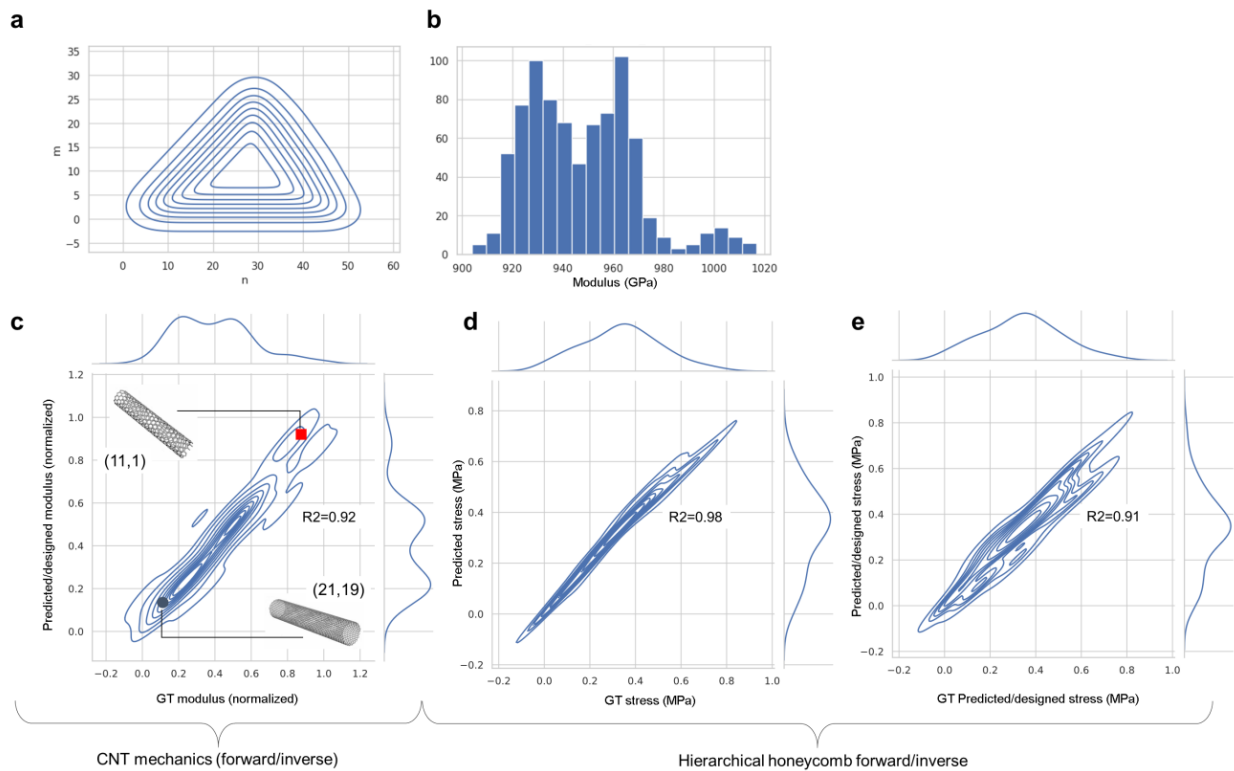

**Figure 11**: Analysis and design of CNTs, solving both forward and inverse problems. Panel a shows the chirality (n,m) design space in the training set, and panel b the distribution of the Young's modulus (in GPa). Panel c shows the results for both forward and inverse problem in one plot, highlighting two designs generated for CNTs with extreme properties. In this plot, the minimum and maximum moduli are normalized by the minimum and maximum moduli in the training set (normalized modulus=0 corresponds to 903.85 GPa, and normalized modulus=1 corresponds to 1,016.95 GPa).

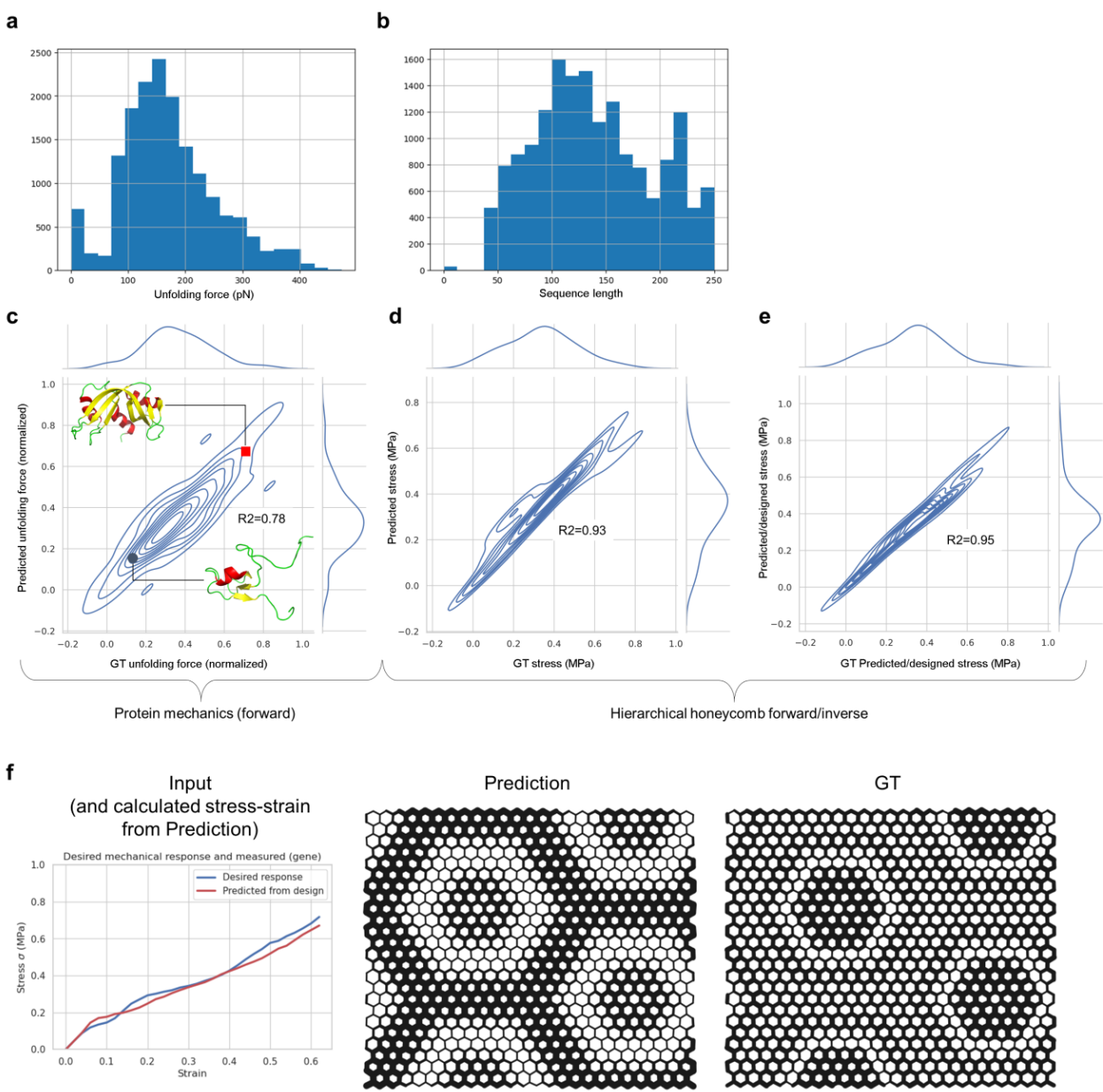

**Figure 12**: Analysis and design of protein unfolding mechanics. Panel a shows the distribution of the maximum force during unfolding (in pN), as well as the sequence length in panel b. Panel c shows the results for the prediction task, identifying the maximum unfolding force from the amino acid sequence. Two proteins with extreme properties are highlighted. In agreement with common understanding, the strongest protein (red square) is highly organized and rich in beta-sheet secondary structure (reaching a strength of ~ 376.68 pN; PDB ID 1YMN). The weaker protein (blue circle) reaches a strength of only 74.1 pN; PDB ID 2DID). The relative weakness of this protein can be explained by the structure, showing more disordered regions, and only a small helical and sheet domain fraction. In this plot, the maximum unfolding force is normalized by the maximum value (normalization factor=475 pN). To show that the model is still capable of solving hierarchical honeycomb design tasks (in alignment with the excellent R2 values seen in panels d and e), we depict a sample design task in panel f.

**Table 1:** Summary of all prompts used in the model (applied to: hierarchical honeycomb materials, carbon nanotubes, and proteins). All tasks take the format `Task<input_to_task>` **`[output]`**. During training, samples of the entire task and output is provided and trained using causal masking. During inference, we provide only the task input and the model then completes the sequence by providing the answer.

| Task input | Description | Output example | |
|---|---|---|---|
| `GetProperties<113, 43, 20,113, 20, 20, 43, 47...>` | Calculate stress-strain response from VQ-VAE codebook language encoding | `[-0.988,-0.914,-0.840,-0.769,-0.702,-0.641,...]` | Forward task |
| `GeneGetProperties< 26, 25, 19, 14, 0, 0, 0, 0, 0, 0>` | Calculate stress-strain response from gene | `[-0.988,-0.914,-0.840,-0.769,-0.702,-0.641,...]` | |
| `GetProteinForce<MSYYHHHHHLESTSL YKKAGSENLYFQGNKKDIPWTDLNRASGVGST GILQARIINGVIYVRGNSIPVPNVAPNFIVPV GTFPPAFGTNLPQFDSSGTFYSHGNLSLSLIN MSPSGIAVGNPNNTSMNGKTISFALSAPLL>` | Calculate unfolding strength of a protein defined by its sequence | `[0.365]` | |
| `GetCNTModulus< 29, 26>` | Calculate Young's modulus of a CNT defined by chirality (*n,m*) | `[0.174]` | |
| `GetStructure<-0.987,-0.915,-0.841,-0.773,-0.709...>` | Design VQ-VAE codebook language encoding to meet stress-strain response | `[113, 43, 20,113, 20, 20, 43, 47, 99,...]` | Inverse task |
| `GeneGetStructure<-0.988,-0.914,-0.840,-0.769...>` | Design gene to meet stress-strain response | `[ 10, 13, 14, 10, 9, 0, 0, 0, 0, 0]` | |
| `GetCNTStructure<0.558>` | Design CNT (*n,m*) with a specific Young's modulus | `[ 44, 0]` | |